\documentclass{nature}

\usepackage{graphicx}
\usepackage{amssymb}
\usepackage[labelfont=bf]{caption}
\usepackage{subcaption}
\usepackage{xcolor}
\usepackage{gensymb}
\usepackage{lineno}

\usepackage[style=nature,autocite = superscript]{biblatex}
\usepackage[margin=1in]{geometry}
\usepackage{amssymb}
\usepackage{sidecap}
\usepackage{amsmath} 

\bibliography{references.bib}

\makeatletter
\let\saved@includegraphics\includegraphics
\AtBeginDocument{\let\includegraphics\saved@includegraphics}
\renewenvironment*{figure}{\@float{figure}}{\end@float}
\makeatother

\title{Reconstructing the polar interface of infinite-layer nickelate thin films}

\author{Berit H. Goodge$^{1,2*}$, 
Benjamin Geisler$^{3}$,
Kyuho Lee$^{4,5}$,
Motoki Osada$^{4,6}$,
Bai Yang Wang$^{4,5}$,
Danfeng Li$^{4,7}$,
Harold Y. Hwang$^{4,8}$,
Rossitza Pentcheva$^{3}$,
Lena F. Kourkoutis$^{1,2*}$
}

\begin{document}

\maketitle
\begin{affiliations}
\item{School of Applied and Engineering Physics, Cornell University, Ithaca, NY 14853, USA} 
\item{Kavli Institute at Cornell for Nanoscale Science, Cornell University, Ithaca, NY 14853, USA} 
\item{Department of Physics and Center for Nanointegration (CENIDE), University of Duisburg-Essen, Lotharstrasse 1, 47057 Duisburg, Germany}
\item{Stanford Institute for Materials and Energy Sciences, SLAC National Accelerator Laboratory, Menlo Park, CA 94025, USA} 
\item{Department of Physics, Stanford University, Stanford, CA 94305, USA} 
\item{Department of Materials Science and Engineering, Stanford University, Stanford, CA 94305, USA} 
\item{Department of Physics, City University of Hong Kong, Kowloon, Hong Kong, China}
\item{Department of Applied Physics, Stanford University, Stanford, CA 94305, USA}
 
\end{affiliations}

\noindent Correspondence to:\\ 
Berit H. Goodge, Lena F. Kourkoutis \\
email: bhg37@cornell.edu, lena.f.kourkoutis@cornell.edu



\newpage
\begin{abstract}
Nickel-based superconductors provide a long-awaited experimental platform to explore possible cuprate-like superconductivity.
Despite similar crystal structure and $d$ electron filling, these systems exhibit several differences. 
Nickelates are the most polar layered oxide superconductor, raising questions about the interface between substrate and thin film -- thus far the only sample geometry to successfully stabilize superconductivity. 
We conduct a detailed experimental and theoretical study of the prototypical interface between Nd$_{1-x}$Sr$_x$NiO$_2$ and SrTiO$_3$.
Atomic-resolution electron energy loss spectroscopy in the scanning transmission electron microscope reveals the formation of a single intermediate Nd(Ti,Ni)O$_3$ layer.
Density functional theory calculations with a Hubbard $U$ term show how the observed structure alleviates the strong polar discontinuity. 
We explore effects of oxygen occupancy, hole doping, and cation structure to disentangle the contributions of each for reducing interface charge density. 
Resolving the nontrivial interface structure will be instructive for future synthesis of nickelate films on other substrates and in vertical heterostructures.
\end{abstract}



\newpage


The stabilization of superconductivity in thin film hole-doped infinite-layer rare-earth nickelates, $R$NiO$_2$ ($R$ = La, Nd, Pr), provides the first experimental platform in which to explore a possible analogue to the high-T$_c$ cuprates \cite{li_superconductivity_2019, zeng2020phase, gu2020single, osada2020superconducting,zeng2021superconductivity,osada2021nickelate, gao2021preparation}.
The availability of real, measurable samples thus demands comparison with theoretical predictions of the past decades \cite{anisimov_electronic_1999, lee_infinite-layer_2004}.
Already, such comparisons have revealed a number of distinctions between the nickelate and cuprate superconductors, including differences in the hybridization, charge-transfer energy, and pairing mechanism \cite{hepting2020electronic, gu2020single, goodge2021doping}. 
On the other hand, theoretical calculations have suggested that depletion of the rare-earth bands could result in a Fermi surface reconstruction that renders the nickelates cuprate-like despite these differences \cite{geisler2020fundamental}. 
A more practical distinction between the nickelate and cuprate systems so far, however, is the geometry of stable, superconducting samples. 
Unlike the bulk- and powder-synthesized cuprate compounds first demonstrated \cite{bednorz_possible_1986,  wu1987superconductivity, maeda1988new,sheng1988bulk,schilling1993superconductivity}, all reports of nickelate superconductivity to date \cite{li_superconductivity_2019, zeng2020phase, gu2020single, osada2020superconducting,zeng2021superconductivity,osada2021nickelate, pan2021superconductivity, ren2021superconductivity} have been in thin-film geometries where the crystalline substrate (and, in some cases, thin capping layer) provide epitaxial support for the infinite-layer phase \cite{lee_aspects_2020}. 
The synthesis of bulk infinite-layer samples has shown promising progress \cite{puphal2021topotactic} but has yet to stabilize superconductivity in this form. 

This specificity has in turn spurred questions about the interface between substrate and nickelate film and the role it plays in the observed superconductivity \cite{geisler2020fundamental, zhang2020similarities,   bernardini2020stability,he2020polarity,geisler2021correlated}.
In particular, a large polar discontinuity should exist at an atomically abrupt, ideal interface between the charge-neutral planes of the substrate (Sr$^{2+}$O$^{2-}$ and Ti$^{4+}$O$^{2-}_2$) and the charged planes of the film ($R^{3+}$ and Ni$^{1+}$O$^{2-}_2$) \cite{geisler2020fundamental,he2020polarity}.
Several studies have investigated the possible effect of such an interface, predicting the formation of a high carrier density two-dimensional electron gas (2DEG) \cite{geisler2020fundamental, zhang2020similarities, geisler2021correlated} or other electronic instability \cite{bernardini2020stability}. 
Some of these predictions hearken back to the discovery of superconductivity at the polar interface between LaAlO$_3$ and SrTiO$_3$ \cite{ohtomo2004high,reyren2007superconducting}, raising the question of whether the observed transport in this new family of superconductors should be associated more closely with the high-T$_{c}$ cuprates or with exotic interface phenomena. 
Furthermore, these theoretical studies have indicated how subtly different interface reconstructions can have drastic effects on the electronic structure \cite{zhang2020similarities, geisler2020fundamental,  bernardini2020stability,he2020polarity,geisler2021correlated}.

An atomic-scale understanding of both the chemical and electronic structure of the substrate-nickelate interface is therefore of great interest to the field. 
Are the interface and surface electronically distinct from the rest of the thin film? 
Is superconductivity in the experimentally available thin films representative of the bulk state?
If the observed superconductivity is significantly impacted by interface effects, will stabilizing the same response ever be possible in bulk-grown samples?
As stated by Botana, et al., ``the first important question to clarify is which interface is actually realized in these systems'' \cite{botana2021nickelate}. 

Here, we used atomically-resolved electron energy loss spectroscopy (EELS) to resolve the atomic-scale lattice and electronic structure of the NdNiO$_2$ - SrTiO$_3$ interface, providing key insights into the role of the interface for the observed superconductivity in infinite-layer nickelates. 
Our measurements show the formation of a single intermediate Nd(Ti,Ni)O$_3$ layer between the substrate and nickelate film, which forms during the perovskite phase growth and persists after the topotactic reduction to the infinite-layer phase. 
Density functional theory calculations with a Hubbard $U$ term (DFT+$U$) show that the experimentally observed interface structure effectively quenches the formation of an interface 2DEG, consistent with the lack of excess interface charge measured experimentally by EELS. 
Systematic variation of the model interface disentangles the contributions from the atomic stacking sequence, oxygen occupancy, and lattice distortions.
Together, our experimental measurements and theoretical calculations provide important insights needed to establish a realistic picture of this novel superconducting system.

\section*{Results}

\begin{figure}
    \centering
        \includegraphics[width=\linewidth]{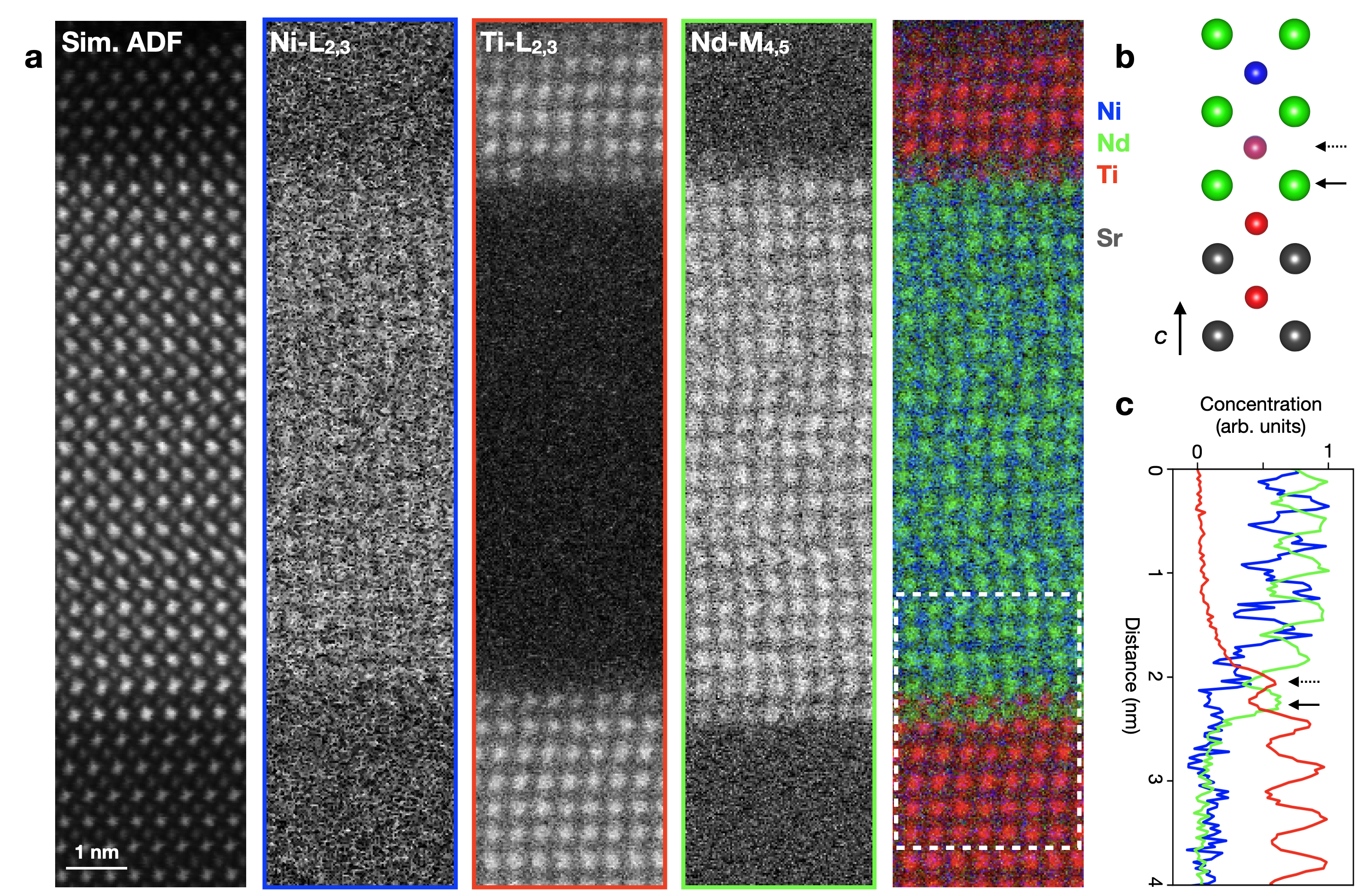}
    \caption{ \textbf{a,} Atomic-resolution elemental mapping by electron energy loss spectroscopy of an NdNiO$_2$ film on SrTiO$_3$ reveals the formation of an intermediate interface titanate layer, drawn schematically in \textbf{b}. \textbf{c,} Concentration profiles of Ni, Ti, and Nd across the interface from the region marked by the dashed white box in \textbf{a} show that the layer is predominantly Ti with some Ni occupancy (dashed arrow). The first Nd-containing atomic plane (solid arrow) forms below the final Ti-containing atomic plane, disrupting the ideal SrTiO$_3$ - NdNiO$_2$ interface.  }\label{fig:map}
\end{figure}

Atomic-resolution elemental mapping by electron energy loss spectroscopy (EELS) in the scanning transmission electron microscope (STEM) across the substrate - nickelate interface reveals the formation of a single intermediate layer of Nd(Ti,Ni)O$_y$ (Figure \ref{fig:map}, $ 2 \leq y \leq 3$). 
An integrated concentration profile across the interface suggests the B-site cation stoichiometry in this layer is predominantly Ti, with some Ni occupancy. 
This intermediate layer forms during the perovskite phase growth, but is retained during the chemical reduction process to the infinite-layer phase. 
Furthermore, it is thus far also observed in all nickelate films stabilized on SrTiO$_3$, regardless of chemical (Sr) doping or rare-earth species (La, Pr, Nd; see Extended Data S1-S4 and \cite{osada2020superconducting, goodge2021doping, osada2021nickelate}). 
The universality of these observations suggest that this interface layer is not an artefact of very specific growth conditions, but rather some inherent mechanism within the system. 
Namely, the atomic interface structure may help alleviate the very strong polar discontinuity described above that would otherwise form at an abrupt interface \cite{nakagawa2006some}. 

\begin{figure}
    \centering
        \includegraphics[width=\linewidth]{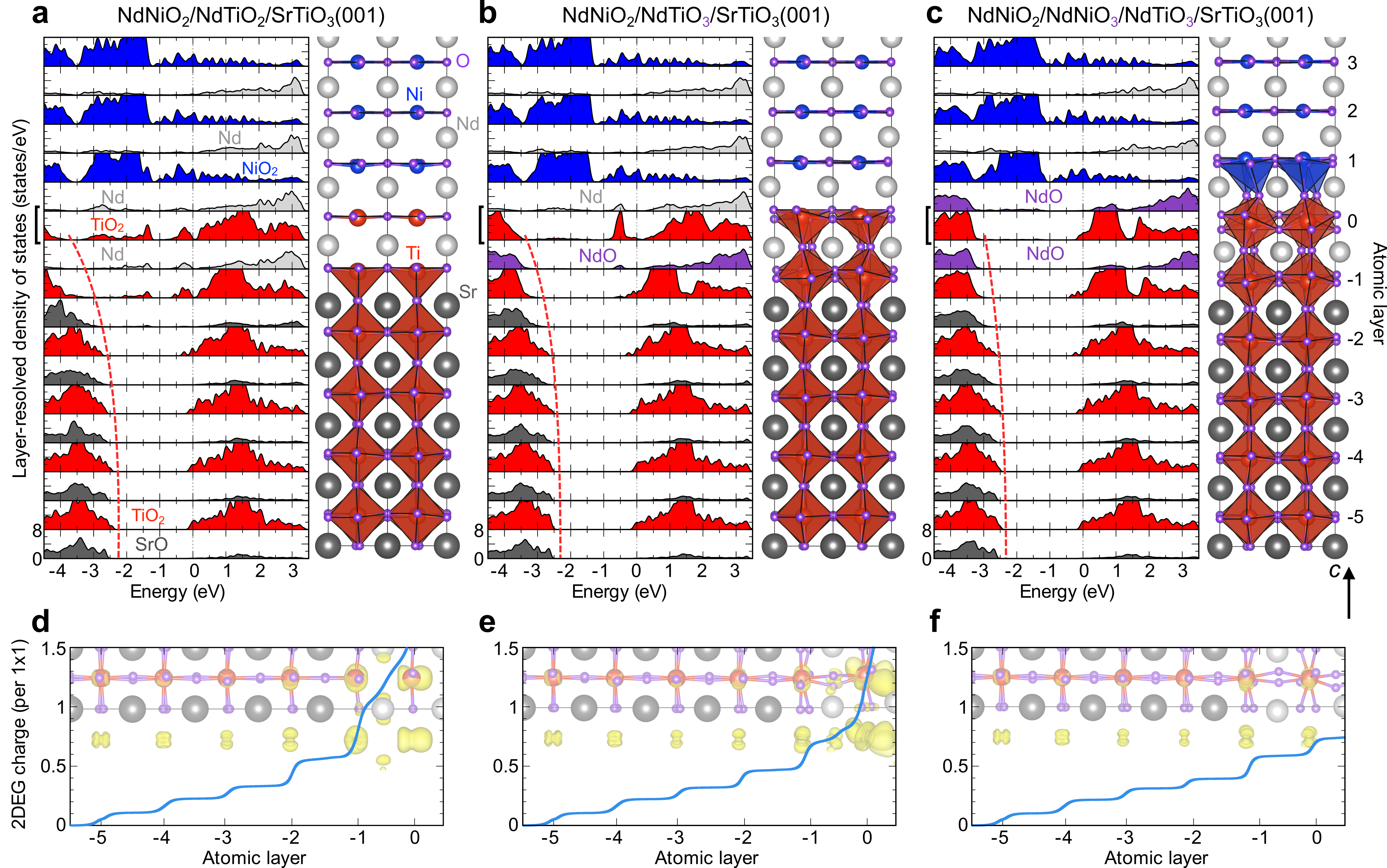}
    \caption{\textbf{a-c, } Optimized atomic geometries and layer-resolved densities of states for three distinct interface structures between NdNiO$_2$ and SrTiO$_3$(001), each of which is consistent with the cation structure observed in Figure 1, assuming a perfect TiO$_2$ layer.  \textbf{d-e, } Visualization of the 2DEG electron density (integrated from -0.7 eV to $E_F$) and calculated accumulated 2DEG charge for the interfaces shown in \textbf{a-c}, respectively.  }\label{fig:2DEG}
\end{figure}

The impact of excess Ti at the interface on potential charge buildup is examined using density functional theory (DFT). 
Previous work has shown that an abrupt interface between SrTiO$_3$ and NdNiO$_2$ would host a high carrier density 2DEG, but that a single intermediate perovskite NdNiO$_3$ layer may be sufficient to prevent the formation of such a 2DEG \cite{geisler2020fundamental}. 
Here, we therefore explore the effect of oxygen occupancy near the experimentally observed interface, taking into account the intermediate layer (Figure 2). 
Note that to systematically disentangle the contributions of different lattice effects, we initially neglect any Ni in this intermediate layer, which we initially model as a perfect TiO$_2$ layer. 
Layer-resolved densities of states are calculated for each atomic plane extending from the SrTiO$_3$ substrate into the NdNiO$_2$ film for three cases of oxygen occupancy: 
an abrupt transition from perovskite to infinite-layer, such that all Nd-containing planes are solely Nd and devoid of oxygen (Figure 2a); 
a single NdO plane between the SrTiO$_3$ and intermediate TiO$_2$ plane such that no Ni-containing planes have perovskite-like octahedral coordination (Figure 2b); 
and a full perovskite NdTiO$_3$ intermediate layer followed by infinite-layer NdNiO$_2$ such that the first Ni-containing plane is partially coordinated (Figure 2c).
Atomic layers are indexed from the interface NdTiO$_y$ layer 0, with each BO$_2$ plane increasing (decreasing) by integer values into the film (substrate). 
The AO and A planes are labelled by the intermediate half-integers. 

The DFT+$U$ calculations demonstrate how the presence of excess interface charge depends strongly on the oxygen stoichiometry of each layer.
The first two cases in particular show pronounced band-bending (dashed red lines) especially in the substrate. 
Each of the structures features a finite occupation of Ti 3$d$ states, particularly $d_{xy}$ orbitals. 
The total charge of the 2DEG, calculated by progressively integrating the states between -0.7 eV and $E_F$, is comparatively smallest for the SrTiO$_3$ - NdTiO$_3$ - NdNiO$_3$ - NdNiO$_2$ interface structure shown in Figure 2c.
In all cases, however, the excess charge is localized and the Ti sites suggest lower Ti valence at the interface which is considerably modified towards $\sim$3+ (rather than 4+). 

While the annular dark-field (ADF) collection geometry used for the left-most STEM image in Figure 1a is sensitive to heavy nuclei (such as the Nd, Sr, Ti, and Ni cations in this system), annular bright-field (ABF)-STEM imaging is required for imaging lighter elements such as oxygen. 
Empty and filled apical oxygen sites in the infinite-layer and perovskite structures in the NdNiO$_2$ film and SrTiO$_3$ substrate respectively are indeed visible on either side of the interface in an atomic-resolution ABF-STEM image (Extended Data S5), but the technique is not suited to the kind of quantitative measurements at the intermediate layers which are required to distinguish between the three models in Figure 2.
To determine the oxygen occupancy and investigate the local Ti valence of a real interface, we again harness the combined high spatial and energy resolution of EELS in the STEM.

\begin{figure}
    \centering
        \includegraphics[width=\linewidth]{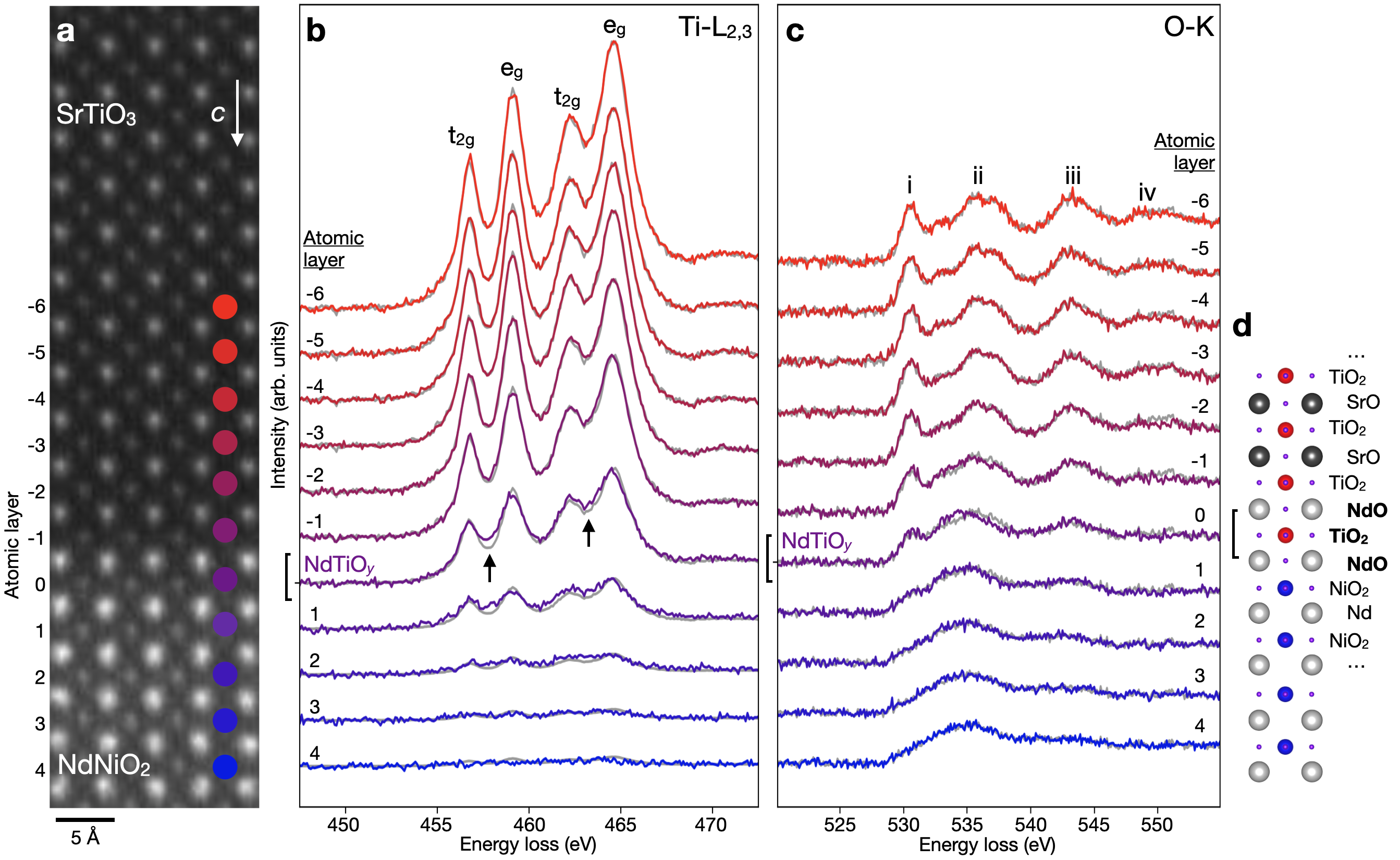}
    \caption{
    \textbf{a,} High-angle annular dark-field (HAADF)-STEM image of the substrate - film interface with the atomic layers labelled according to the conventions used here. The growth direction is oriented downwards as indicated by the arrow $c$. Colored dots indicate the atomic layers for which each spectrum is a representative measurement (see Methods). \textbf{b,} The Ti-L$_{2,3}$ edge shows a minor filling between the $e_g$ and $t_{2g}$ peaks at the intermediate layer 0. \textbf{c,} The O-K edge reflects the progression from bulk-like SrTiO$_3$ to NdNiO$_2$. Also plotted are best-fit linear combinations of SrTiO$_3$ and NdNiO$_2$ references (grey). \textbf{d,} Schematic representation of the observed interface structure with the atomic planes labeled. }\label{fig:EELS}
\end{figure}

In addition to elemental mapping, STEM-EELS provides the capability to measure local charge modulations and interface effects \cite{ohtomo2002artificial, mundy2014visualizing}. 
Figure 3 shows atomically resolved high energy resolution STEM-EEL spectra which probe the electronic states across the SrTiO$_3$ - NdNiO$_2$ interface. 
We note that for clarity of the plots, the vertical orientation of the interface has been inverted from Figure 1, such that the SrTiO$_3$ substrate is on top and the growth direction ($c$-axis) points down (Figure 3a). 
Summed and background-subtracted spectra showing the energy loss near-edge structure (ELNES) of the Ti-L$_{2,3}$ and O-K EELS edges of each atomic layer are plotted in Figure 3b and 3c, respectively. 
Each spectrum is plotted over bulk references shown in grey, which are acquired during the same measurements from regions more than 10 unit cells from the interface. 
The Ti-L$_{2,3}$ spectra are referenced against measurements from the SrTiO$_3$ substrate. 
The O-K spectra at each layer are compared to a best-fit linear combination of bulk-like references from SrTiO$_3$ and NdNiO$_2$ as determined by a least squares minimization $C_1 \, \textrm{O}_{STO} + C_2 \, \textrm{O}_{NNO}$, where $C_1$ and $C_2$ are non-negative and O$_{STO}$ and O$_{NNO}$ are the respective reference spectra.
More details of the minimization are given in Extended Data S6 and S7. 
The end-member spectra for SrTiO$_3$ and NdNiO$_2$ are consistent with `bulk-like' measurements of both compounds reported previously \cite{hepting2020electronic, goodge2021doping}.
Spectra in the film are consistent with full oxygen reduction to the infinite-layer phase \cite{goodge2021doping}, while the substrate layers show no sign of induced oxygen vacancies \cite{zhou2021negligible}, consistent with the high reduction energy of SrTiO$_3$ \cite{geisler2020fundamental, sahinovic2021active}.

Intriguingly, the Ti-L$_{2,3}$ spectra show very little modification across the interface, other than a reduction in overall intensity consistent with the observed stoichiometry (Figure 1) and differences in the orbital bonding environment indicated by a subtle change to the L$_{2,3}$ edge at the interface layer 0, marked by the black arrows.
There is no sign of significant Ti$^{3+}$ \cite{ohtomo2002artificial} at the interface layer, as predicted by the theoretical NdTiO$_3$ model in Figure 2c. 
Instead, the small change in the L$_{2,3}$ edge is likely more reflective of subtle changes to the coordination environment in the interface layer induced by the change in A-site cation from Sr to Nd and the partial Ni occupancy in this layer so far neglected in our models.  

By comparison, the O-K edge,  which can be used to track chemical and electronic states as well as bonding environments in complex oxides \cite{de_groot_oxygen_1989,kourkoutis_atomic-resolution_2010}, shows a more significant response near the interface. 
In layers -1 to 1, the O-K edge spectra show clear deviations from reference linear combinations of the two end-members SrTiO$_3$ and NdNiO$_2$. 
At atomic layer -1, the first response observed in the O-K ELNES is a broadening of the `ii' peak, which tracks the A-site cation in complex oxides \cite{kourkoutis_atomic-resolution_2010}, reflecting the proximity of this layer to both Sr and Nd cations.
The intermediate layer, atomic layer 0, is fully surrounded by Nd-containing layers and shows a more complete shift of this peak into the Nd position. 
Layer 1 can be almost entirely described by the NdNiO$_2$ reference (Extended Data S6 and S7), with only a small residual of peak `i' remaining, possibly due to contributions from probe tails extending into layer 0.
The stoichiometry of layer 1 suggests that the B-sites are essentially fully occupied by Ni, with very little or no Ti extending to this layer (see Figure 1a,c). 
Layers 2 and above appear to be essentially fully consistent with a fully-reduced infinite-layer phase and Ni 3$d^9$ states \cite{goodge2021doping}.

\begin{figure}
    \centering
        \includegraphics[width=0.5\linewidth]{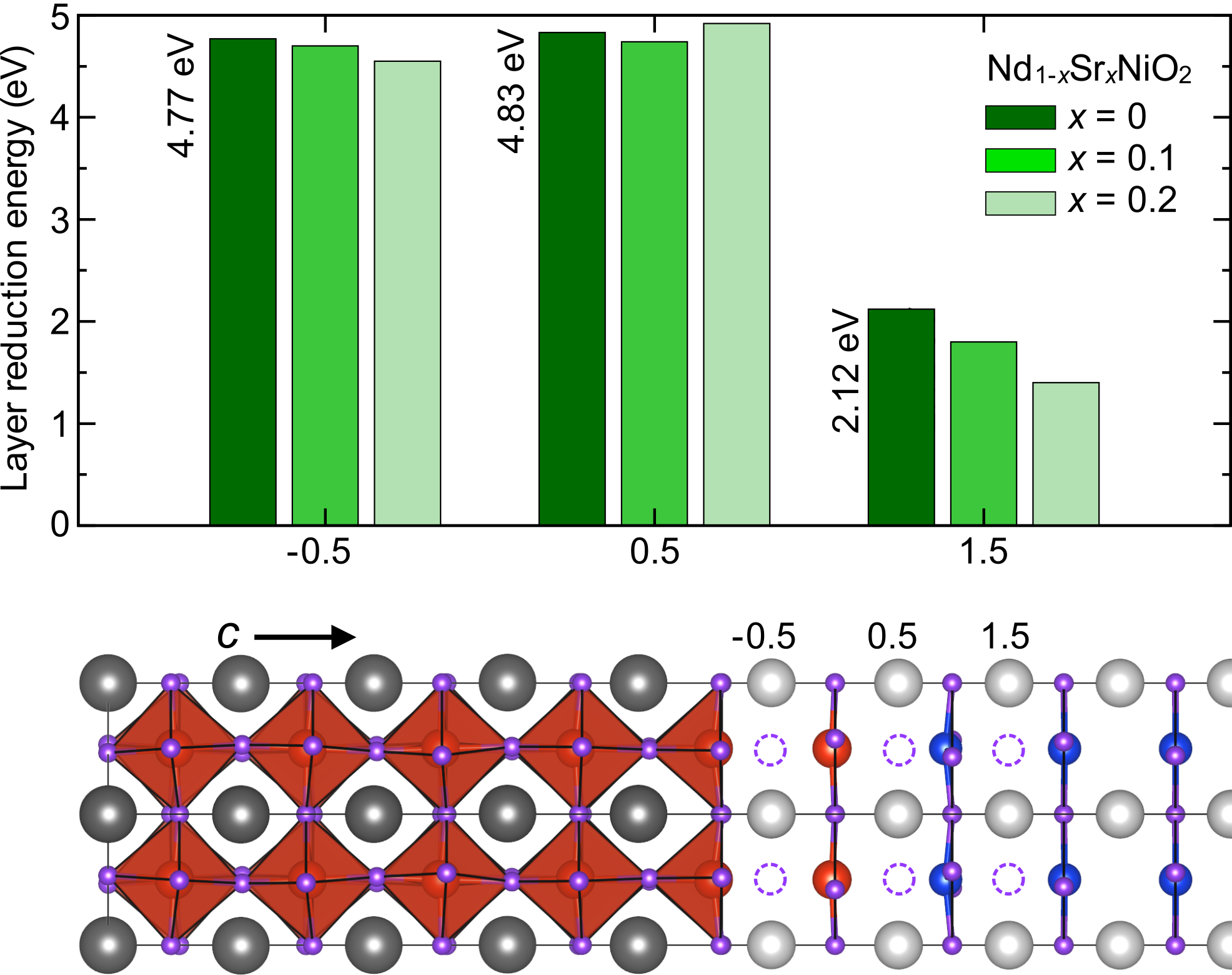}
    \caption{Calculated energy required to remove apical oxygens in the film cation layers near the interface for different levels of Sr doping in the Nd$_{1-x}$Sr$_x$NiO$_2$ film, including undoped, lightly-doped $x$ = 0.10, and optimally doped $x$ = 0.20. }
    \label{fig:reduction}
\end{figure}

DFT+$U$ calculations validate this experimentally-derived structure by calculating the per-layer reduction energy required to remove apical oxygens from the perovskite phase of each layer in the interface. 
Figure 4 shows that the reduction energy near Ti is significantly higher than near Ni, qualitatively similar to previous calculations on related atomic interfaces \cite{geisler2020fundamental} . 
Thus, the finite concentration of Ti in the intermediate layer impedes the oxygen reduction from the neighboring AO planes.
The effect of Sr doping in the nickelate film, on the other hand, is found to have only a modest impact on the reduction energy.
We therefore expect that the overall atomic structure of the substrate-film interface will be as described here for any chemical doping realized thus far \cite{li2020superconducting,osada2020superconducting,osada2021nickelate}, while experimental reports of small differences in the optimal reduction conditions for each case may reflect the subtle details revealed by our calculations. 
Together, the chemical fingerprinting and theoretical calculations described here allow us to fully reconstruct the atomic interface stoichiometry -- including cation as well as oxygen occupancy -- pictured schematically in Figure 3d.

\begin{figure}
    \centering
        \includegraphics[width=\linewidth]{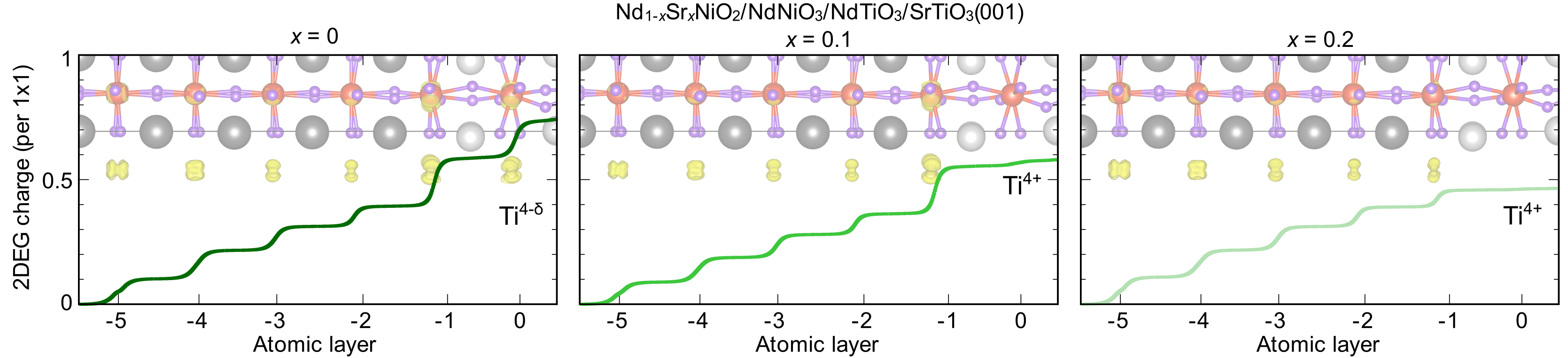}
    \caption{The impact on accumulated interface charge in Nd$_{1-x}$Sr$_x$NiO$_2$ films by systematic variation of Sr-doping for (left to right) undoped, lightly-doped $x$ = 0.10, and optimally doped $x$ = 0.20, assuming perfect TiO$_2$ stoichiometry in the intermediate layer (layer 0, see Figure 2c).  }\label{fig:doping}
\end{figure}

\begin{figure}
    \centering
        \includegraphics[width=0.35\linewidth]{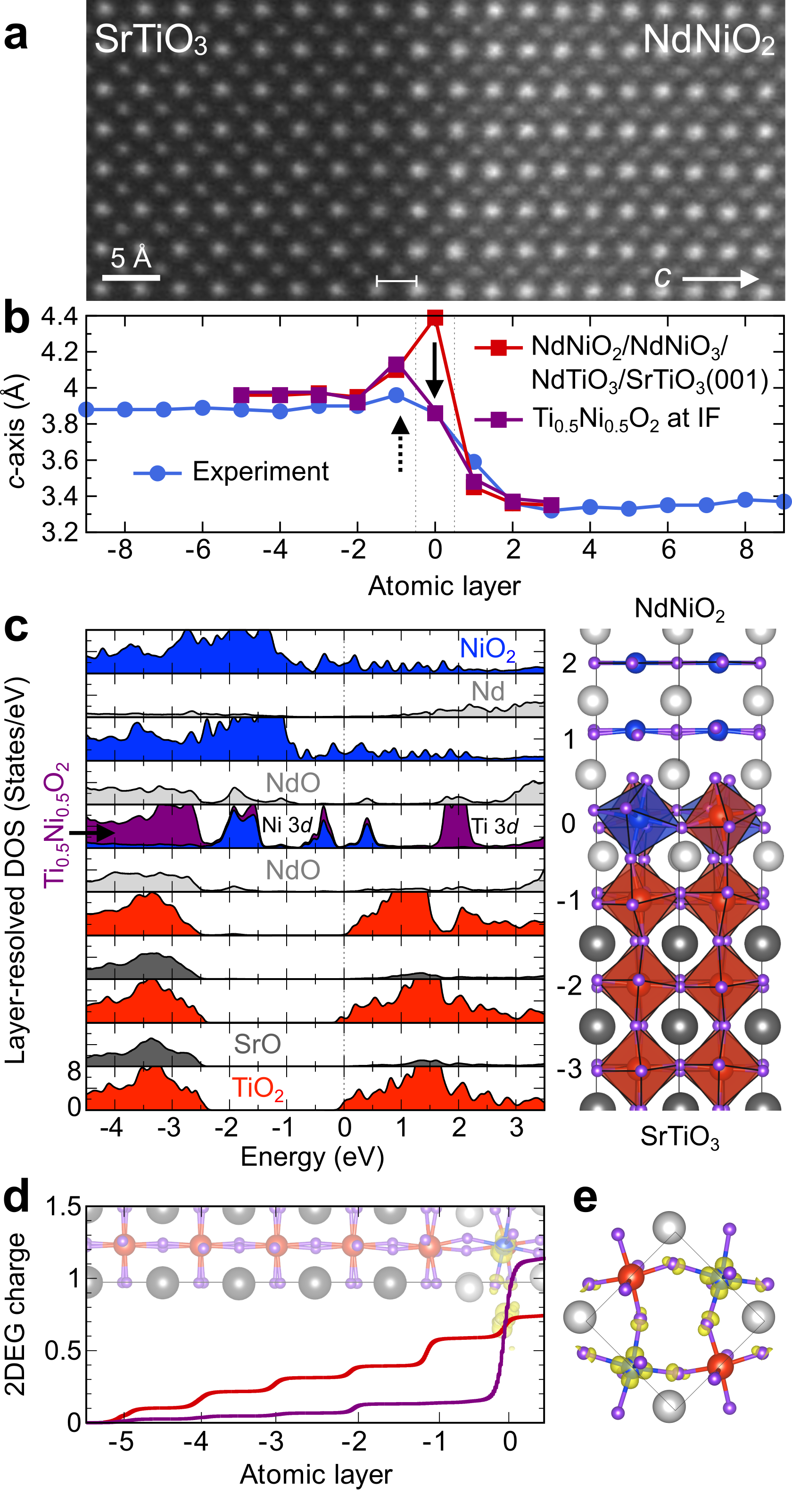}
    \caption{\textbf{a,} HAADF-STEM image of a real interface used to measure the layer-resolved local $c$-axis lattice distance between rare-earth planes (e.g., SrO - SrO or Nd - Nd). \textbf{b,} Comparison between experimental measurements (blue) and DFT+$U$ results for a pure NdTiO$_3$ intermediate layer 0 (red, same as Figure 2c) and 50\% Ni substitution into the intermediate layer (purple). \textbf{c,} Layer-resolved densities of states for the interface model with a mixed NdTi$_{0.5}$Ni$_{0.5}$O$_3$ interface layer. The purple states at layer 0 represent Ti$_{0.5}$Ni$_{0.5}$O$_2$. The blue states in this layer are Ni 3$d$ contributions. The states at $\sim$2 eV are purely of Ti-3$d$ character. \textbf{d,} Accumulated 2DEG charge for the pure and mixed intermediate layer models. \textbf{e,} Jahn-Teller distortions in the mixed intermediate layer ``absorb'' excess charge at the interface in the Ni 3$d_{x^2-y^2}$ states, thereby quenching the 2DEG in the SrTiO$_3$ substrate.  \label{fig:lattice}}
\end{figure}

With this more precise atomic picture of the interface, we revisit the model presented in Figure 2c, which is the most representative of the observed cation and oxygen structure, yet still fails to adequately reproduce the observed electronic structure. 
To reconcile the remaining discrepancy between the model -- which suggests a small but finite density of excess charge at the interface -- with the experimental observations -- which show no sign of reduction in Ti valence -- it is necessary to consider and account for the remaining parameter space in this system: namely the intentional chemical (hole) doping of the films and the observed Ni occupancy at the interface layer. 
Figure 5 presents the results of DFT+$U$ calculations for different levels of Sr doping in the NdNiO$_2$ film, using the interface described above. 
Sr doping is used to introduce holes into the film, but is also found to further reduce the interface 2DEG, as observed by the decreasing levels of excess charge from the undoped ($x$ = 0\%) to optimally-doped ($x$ = 20\%) models. 
Even accounting for this chemical doping, however, the system still shows a clear tendency towards some 2DEG formation.

Finally, we consider in conjunction the role of Ni partial occupancy in the interface layer and more subtle structural distortions which may play a role in accommodating the excess charge. 
Quantitative high-angle annular dark-field (HAADF)-STEM measurements can be used to track the $c$-axis lattice spacing in each unit cell across the interface (Figure 6a). 
Similar to the layer-resolved EELS fine structure, in which both the substrate and film exhibit ``bulk-like'' character several unit cells away from the interface, the measured lattice constant of the NdNiO$_2$ film away from the interface is consistent with previous results for the undoped infinite-layer under compressive strain \cite{li2020superconducting}. 
Figure 6b tracks the progression of $c$-axis lattice constant as measured between consecutive A-site planes observed experimentally and for two theoretical models.
In the experimental measurements, the most notable feature is a local expansion of the lattice between the final SrO and the first NdO plane (across layer -1 following the convention used above). 
Structural relaxations of the SrTiO$_3$ - NdTiO$_3$ - NdNiO$_3$ - NdNiO$_2$ interface obtained from DFT+$U$ calculations also show a comparable expansion of the -1 layer, consistent with the experimental observations. 
In this structural model, however, the intermediate layer 0 shows an even larger $c$-axis expansion, which is not observed in the experimentally measured structure.  
Instead, by including Ni in the modeled intermediate layer, very close agreement is achieved between the experimental and theoretical structures. 
Independent of a precise quantification of the exact Ni/Ti ratio at the interface, the inclusion of Ni in this layer dramatically reduces the Nd-Nd distance (far beyond the Ti-Ni ionic radius difference), more closely reproducing the structure observed by HAADF-STEM imaging. 
Electronically, this partial occupancy almost entirely quenches the 2DEG. 
Figure 6e reveals that the excess charge is accommodated exclusively by the Ni-3$d_{x^2 - y^2}$ orbitals via Jahn-Teller distortions which serve as a ``sponge'' to accommodate excess charge and preserve Ti$^{4+}$, as reflected in the projected densities of states shown in Figure 6c and comparison of the integrated 2DEG charges in Figure 6d.
Despite this screening of the polar interface, the Nd-5$d$ states are still depleted in the layers which are included in the model (Figure 6c).

\section*{Discussion}

The combined results of our experimental and theoretical work presented here paint a clear picture of the SrTiO$_3$ - NdNiO$_2$ interface which has so far been the subject of much speculation. 
Rather than a strongly polar abrupt atomic interface hosting a large 2DEG, we show that the combined effects of atomic cation arrangement, subtle structural distortions, and oxygen stoichiometry all but entirely suppress the build-up of any excess charge. 
While these results demonstrate that the substrate-film interface does not itself host superconductivity, DFT+$U$ results show reduced Nd 5$d$ states within at least the first few atomic layers of the film which indicate electronic reconstruction towards a more cuprate-like Fermi surface \cite{geisler2020fundamental, geisler2021correlated}. 
Similar Fermi surface reconstruction is not predicted for bulk NdNiO$_2$ modeled at the same lattice strain provided by the SrTiO$_3$ substrate \cite{geisler2020fundamental}, so the importance of the atomic and electronic structure of the interface studied here for superconductivity can not yet be ruled out.
Future studies should investigate how far into the film (away from the interface) this Nd-5$d$ depletion persists and compare that distance to experimental measurements of the superconducting thickness in films \cite{kozuka2009two, wang2021isotropic}.

Two other possibilities may contribute to the lack of superconducting bulk samples: 1) the substrate provides important structural support during the chemical reduction process, or 2) some degree of lattice strain is necessary to stabilize superconductivity. 
In the first case, it may be possible to explore other routes of oxygen reduction which will not require the mechanical support of a substrate. 
An extreme example of the second case has been demonstrated previously in rutile RuO$_2$, an otherwise metallic compound, which can be made superconducting through very high epitaxial strain imposed by certain substrate orientations \cite{ruf2021strain}. 
Applied and epitaxial strain have also been shown to enhance $T_\textrm{c}$ in superconducting Sr$_2$RuO$_4$ \cite{ hicks_strong_2014, steppke_strong_2017, nair_demystifying_2018}. 
Indeed, a recent report has shown enhancement of $T_\textrm{c}$ by $\sim 70\%$ in Pr$_{0.82}$Sr$_{0.18}$NiO$_2$ thin films on SrTiO$_3$ under pressures beyond 12 GPa \cite{wang2021pressure}.
It seems important to note, however, that in the thin film nickelates the native epitaxial strains are somewhat lower ($0.4\%$ compressive for Nd$_{0.8}$Sr$_{0.2}$NiO$_2$ on SrTiO$_3$ versus anisotropic $2.3\%$ tensile and $4.1\%$ compressive strain for RuO$_2$ and up to $0.6\%$ compressive strain in Sr$_2$RuO$_4$) and the transition temperatures appreciably higher ($\sim5-15$ K for nickelates, $\sim2$ K for RuO$_2$, and up to $\sim3.5$ K for Sr$_2$RuO$_4$). 
It will therefore be instructive to more thoroughly investigate the impact of strain on the nickelate system, either through epitaxial strain or through the application of external strain in uniaxial or biaxial geometries.
Some strain dependence may already by inferred through the differences in La, Pr, and Nd thin films on SrTiO$_3$ \cite{li2020superconducting,osada2020superconducting,osada2021nickelate}, but epitaxial growth on different substrates will offer a way to access more significant strain variation.

In this work, we have identified a number of distinct physical mechanisms at work near the polar SrTiO$_3$ - NdNiO$_2$ interface, including both A- and B-site cation placements, lattice distortions, and chemical doping. 
Guided by experimental STEM-EELS measurements of real samples, a systematic exploration of model systems by DFT+$U$ calculations allows us to disentangle the individual contributions of each variable in the wide parameter space.
Our combined experimental and theoretical results provide a number of insights for establishing a realistic picture of the infinite-layer nickelate superconductors. 
Firstly, despite the strong polar differences of the substrate and film, we have shown that the formation of a single atomic layer of Nd(Ti,Ni)O$_3$ prevents any significant 2DEG which may impact superconductivity or other macroscopic properties observed in these samples by acting as an additional channel for spurious superconductivity.
As superconducting films are realized on other substrates \cite{ren2021superconductivity}, it will also be the subject of future work to explore the different characteristic interfaces that may form in each case.  
More broadly, the complex atomic-scale structure of this interface should be instructive for future efforts to design nanoscale or few-layer heterostructures \cite{ortiz2021superlattice} and create predictive models of similar interfaces.

\section*{Methods}

\subsection{Sample growth and characterization}
The infinite-layer NdNiO$_2$ on SrTiO$_3$ film presented here was synthesized under the `high-fluence’ conditions using precise laser imaging conditions described elsewhere \cite{lee_aspects_2020}. A SrTiO$_3$ epitaxial capping layer was subsequently deposited using conditions described previously \cite{li_superconductivity_2019}. After the growth, an annealing-based topochemical reduction was then employed to achieve the infinite-layer phase using CaH$_2$ as reducing reagent under previously-determined optimal conditions \cite{lee_aspects_2020}.

\subsection{Atomic-resolution STEM, EELS}
Electron transparent TEM samples of each film were prepared on a Thermo Fisher Scientific Helios G4 UX focussed ion beam (FIB) using the standard liftout method. Samples were thinned to $<$30 nm with 2 kV Ga ions, followed by a final polish at 1 kV to reduce effects of surface damage. All specimens were stored in vacuum to prevent possible degradation in air. 

High-angle annular dark-field (HAADF) scanning transmission electron microscopy (STEM) was performed on an aberration-correction FEI Titan Themis at an accelerating voltage of 300 kV with a convergence angle of 30 mrad and inner and outer collection angles of 68 and 340 mrad, respectively. Elemental maps were recorded with a 965 GIF Quantum ER spectrometer and a Gatan K2 Summit direct electron detector operated in counting mode. 

Layer-resolved EELS across the substrate-film interface (Figure 3) was performed with a $<$15 pA probe current on a Nion UltraSTEM operated at 100 kV equipped with a high stability stage, an Enfinium ER spectrometer, and Quefina 2 camera. The effective energy resolution measured by the full width at half-maximum (FWHM) of the zero loss peak (ZLP) was $\sim$0.39 eV. Spectra were recorded as line profiles across the interface extending from the SrTiO$_3$ substrate into the NdNiO$_2$ film with sampling density on the order of $\sim$1 pm per spectrum. Integrating over $\sim$0.2 Å centered on each B-site yielded layer-resolved spectra for each data set, several of which have been summed to produce the high signal-to-noise series presented here. All spectra have been normalized by the total contributed exposure time for each. 

\subsection{Density functional theory}
We performed first-principles calculations in the framework of density functional theory \cite{kohn1965self} (DFT) as implemented in the Quantum ESPRESSO code \cite{giannozzi2009quantum}. 
The generalized gradient approximation was used for the exchange-correlation functional as parametrized by Perdew, Burke, and Ernzerhof \cite{perdew1996generalized}. 
Static correlation effects were considered within the DFT+$U$ formalism \cite{anisimov1993density, cococcioni2005linear} employing $U$ = 4 eV on Ni and Ti sites, in line with previous work \cite{liu2013heterointerface, botana_similarities_2020, geisler2020fundamental, geisler2021correlated}.

Wave functions and density were expanded into plane waves up to cutoff energies of 45 and 350 Ry, respectively. 
Ultrasoft pseudopotentials \cite{vanderbilt1990soft} were used in conjunction with projector augmented wave datasets \cite{blochl1994projector}. 
The Nd-4$f$ electrons were consistently frozen in the core \cite{liu2013heterointerface,geisler2020fundamental, geisler2021correlated, nomura_formation_2019, lechermann2020late, sahinovic2021active}. 
We used a 12 $\times$ 12 $\times$ 1 Monkhorst-Pack $\vec{k}$-point grid \cite{monkhorst1976special} and 5 mRy Methfessel-Paxton smearing \cite{methfessel1989high} to sample the Brillouin zone. 
The ionic positions were accurately optimized, reducing ionic forces below 1 mRy/a.u.

We modeled different infinite-layer nickelate/SrTiO$_3$(001) interfaces with varying stoichiometries in superlattice geometry by using $2a \times 2a \times c$ supercells with two transition metal sites per layer to account for octahedral rotations, strained to the SrTiO$_3$ substrate lattice parameter $a$ = 3.905 \AA. 
The symmetric supercells feature two identical interfaces on each side and comprise 18 monolayers (MLs) in total: 9.5 MLs of SrTiO$_3$ substrate, 1 + 1 MLs of NdTiO$_n$ ($n$ = 2, 3) or mixed Nd(Ti$_{0.5}$Ni$_{0.5}$)O$_3$, 1 + 1 MLs of NdNiO$_n$,and additional 4.5 MLs of NdNiO$_2$. 
The figures show half of each supercell.
The supercell length $c$ was optimized in each case. 
Hole doping was treated explicitly by substituting 10 and 20\% of the Nd sites by Sr.

The layer-by-layer reduction energies \cite{geisler2020fundamental} shown in Figure 4 were calculated from the DFT+$U$ total energies via 
$E_f^i = E_{\textrm{layers}\geq i\, \textrm{reduced}} - E_{\textrm{layers}\geq i+1\, \textrm{reduced}} + \mu_O$, where $\mu_O = \frac{1}{2} E_{O_2}$ models the oxygen-rich limit. 
Here, $i$ = -0.5, 0.5, 1.5 enumerates the A-site layers from the SrTiO$_3$ substrate into the nickelate film.

\section*{Acknowledgements}
B.H.G. and L.F.K. acknowledge support by the Department of Defense Air Force Office of Scientific Research (No. FA 9550-16-1-0305) and the Packard Foundation. This work made use of the Cornell Center for Materials Research (CCMR) Shared Facilities, which are supported through the NSF MRSEC Program (No. DMR-1719875). The FEI Titan Themis 300 was acquired through No. NSF-MRI-1429155, with additional support from Cornell University, the Weill Institute, and the Kavli Institute at Cornell. The Thermo Fisher Helios G4 UX FIB was acquired with support from the National Science Foundation Platform for Accelerated Realization, Analysis, and Discovery of Interface Materials (PARADIM) under Cooperative Agreement No. DMR-1539918. B.G. and R.P. acknowledge support from the German Research Foundation (DFG) within CRC/TRR 80 (project 107745057) Projects G3, G8 and computational time at magnitUDE, granted by the Center for Computational Sciences and Simulation of the University of Duisburg-Essen (DFG grant nos. INST 20876/209-1 FUGG). The work at SLAC/Stanford is supported by the US Department of Energy, Office of Basic Energy Sciences, Division of Materials Sciences and Engineering, under contract number DE-AC02-76SF00515; and the Gordon and Betty Moore Foundation’s Emergent Phenomena in Quantum Systems Initiative through grant number GBMF9072 (synthesis equipment). M.O. acknowledges partial financial support from the Takenaka Scholarship Foundation.

\section*{Author Contributions}
B.H.G. and L.F.K. conceived of the project. L.F.K., R.P., and H.Y.H supervised the research. B.H.G. and L.F.K. performed the electron microscopy, electron energy loss spectroscopy, and corresponding data analysis. B.G. and R.P. performed the theoretical calculations and corresponding analysis. D.L. and M.O. grew and reduced the nickelate films. K.L., D.L., M.O., and B.Y.W. conducted materials and structural characterization. B.H.G., L.F.K., B.G., and R.P. wrote the manuscript. All authors discussed the results and revised the manuscript.

\section*{Competing interests}
The authors declare they have no competing interests.


\printbibliography
\end{document}


\maketitle

\begin{affiliations}
\item{School of Applied and Engineering Physics, Cornell University, Ithaca, NY 14853, USA} 
\item{Kavli Institute at Cornell for Nanoscale Science, Cornell University, Ithaca, NY 14853, USA} 
\item{Department of Physics and Center for Nanointegration (CENIDE), University of Duisburg-Essen, Lotharstrasse 1, 47057 Duisburg, Germany}
\item{Stanford Institute for Materials and Energy Sciences, SLAC National Accelerator Laboratory, Menlo Park, CA 94025, USA} 
\item{Department of Physics, Stanford University, Stanford, CA 94305, USA} 
\item{Department of Materials Science and Engineering, Stanford University, Stanford, CA 94305, USA} 
\item{Geballe Laboratory for Advanced Materials, Stanford University, Stanford, California 94305, USA}
\item{Department of Physics, City University of Hong Kong, Kowloon, Hong Kong, China}
\item{Department of Applied Physics, Stanford University, Stanford, CA 94305, USA}
 
\end{affiliations}

\noindent Correspondence to:\\  
Berit H. Goodge, Lena F. Kourkoutis \\
email: bhg37@cornell.edu, lena.f.kourkoutis@cornell.edu


\newpage

\begin{figure}
    \centering
        \includegraphics[width=\linewidth]{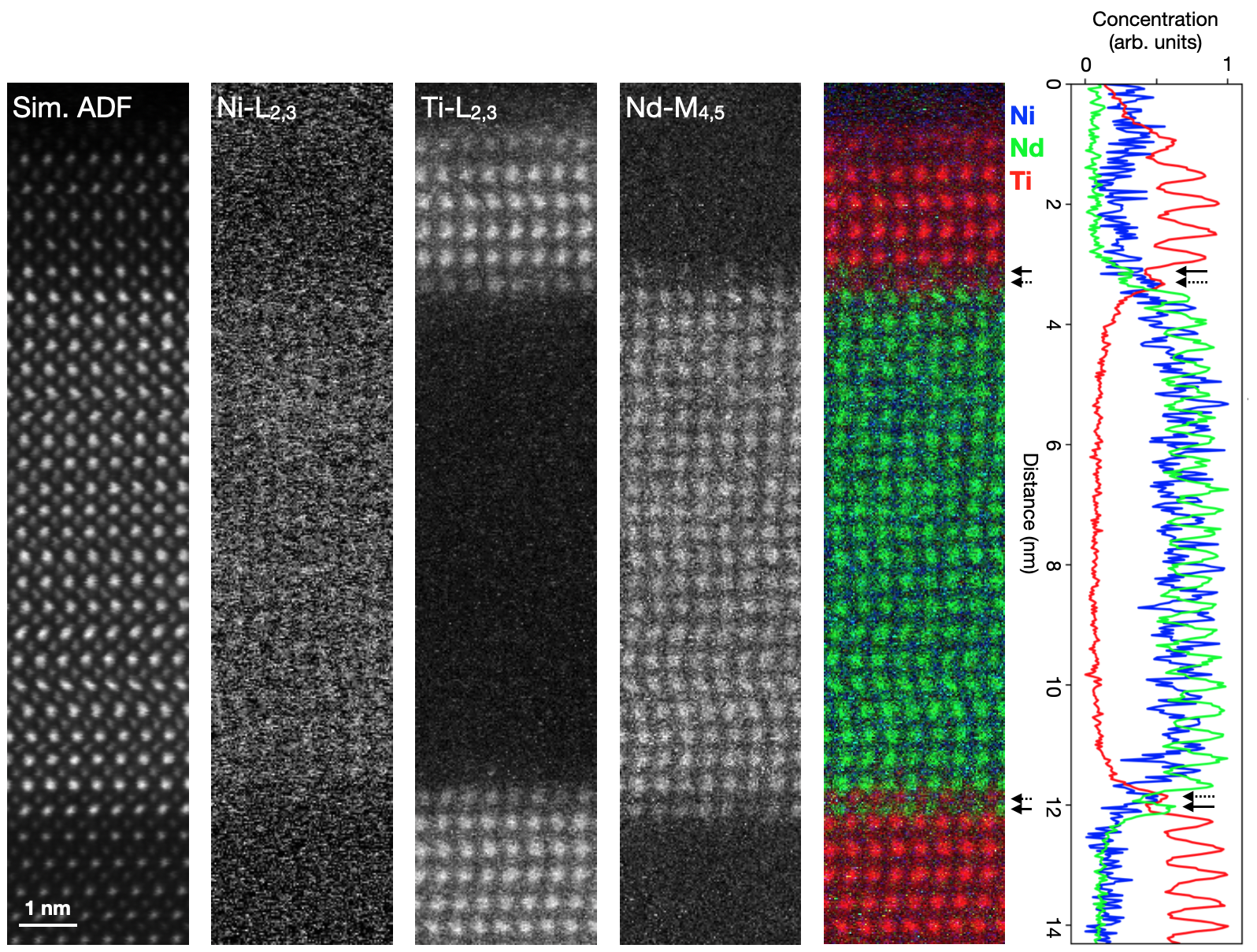}
    \caption{Elemental mapping by STEM-EELS shows the formation of the same Nd(Ti,Ni)O$_3$ intermediate layer at the top interface between film (here, unreduced Nd$_{0.8}$Sr$_{0.2}$NiO$_3$) and SrTiO$_3$ capping layer. }
\end{figure}

\begin{figure}
    \centering
        \includegraphics[width=0.5\linewidth]{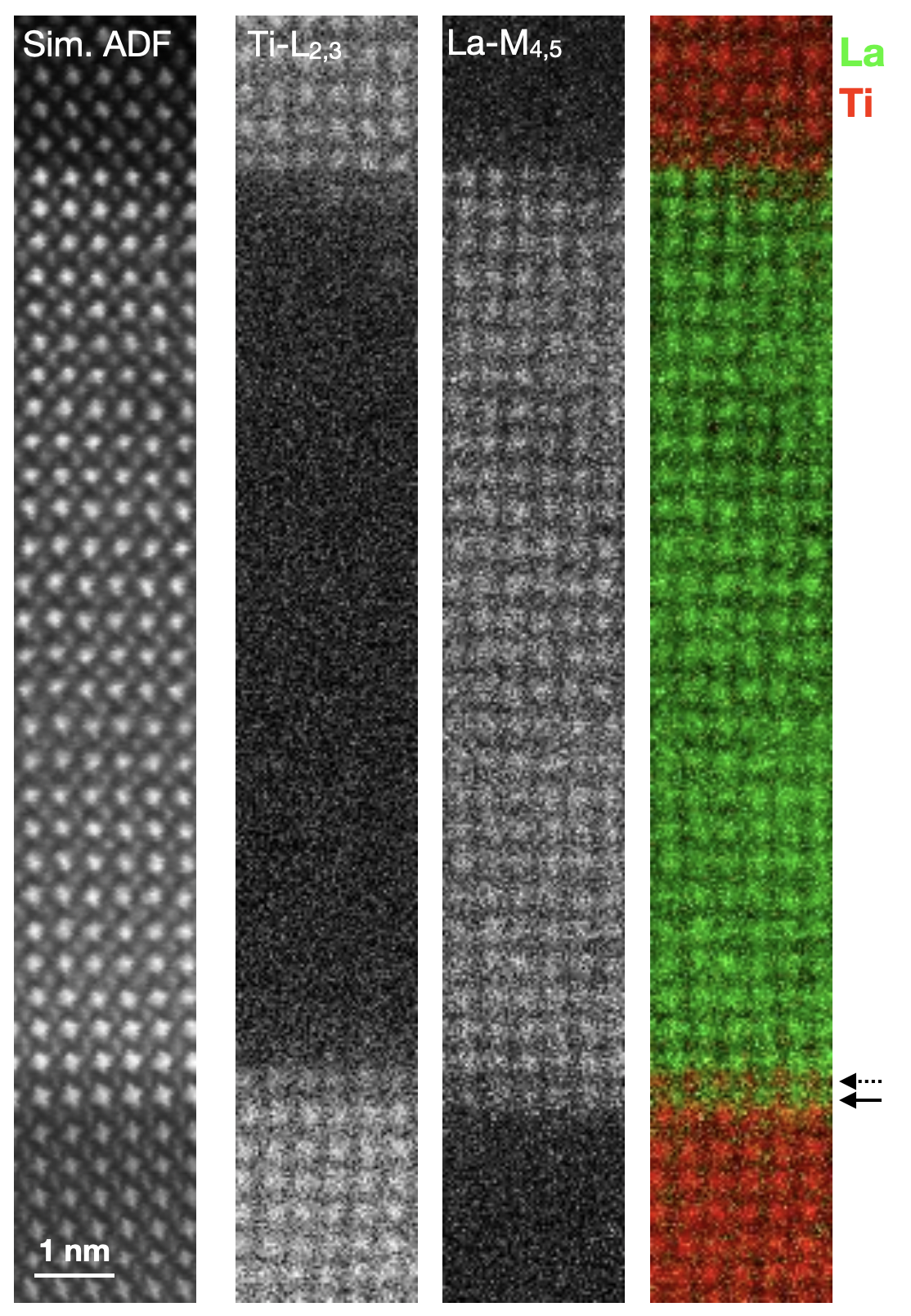}
    \caption{Elemental mapping by STEM-EELS shows the formation of a similar intermediate layer between a LaNiO$_2$ film \cite{osada2021nickelate} and SrTiO$_3$ substrate. }
\end{figure}

\begin{figure}
    \centering
        \includegraphics[width=0.5\linewidth]{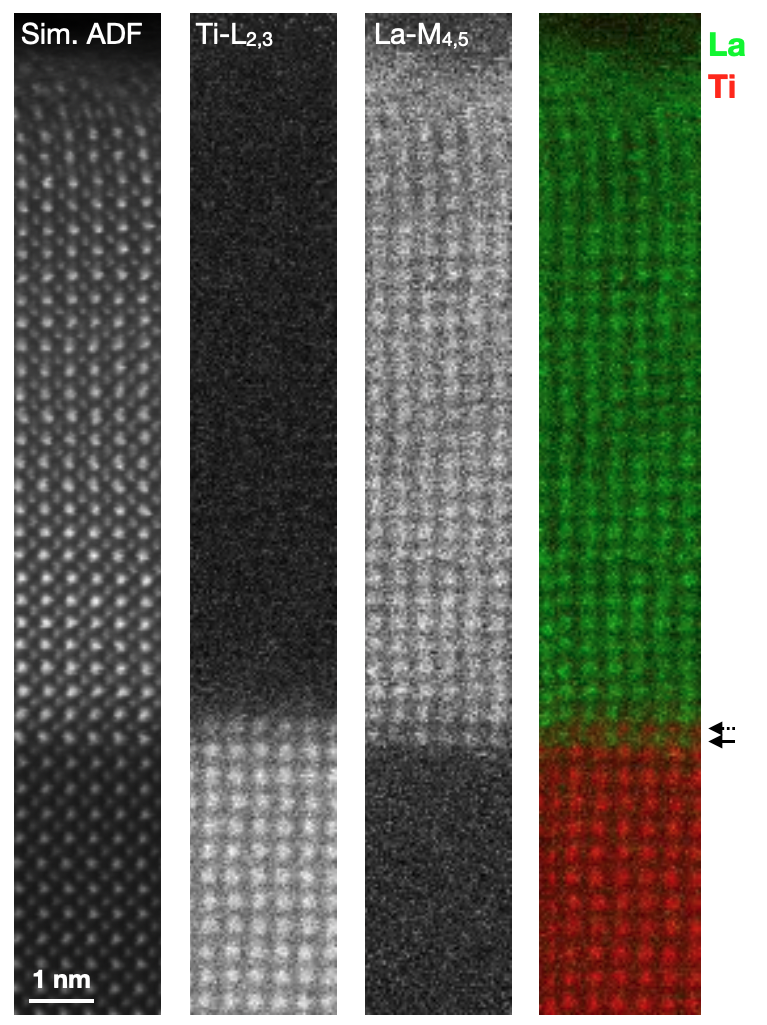}
    \caption{Elemental mapping by STEM-EELS shows the formation of a similar intermediate layer between a La$_{0.8}$Sr$_{0.2}$NiO$_2$ film \cite{osada2021nickelate} and SrTiO$_3$ substrate.}
\end{figure}

\begin{figure}
    \centering
        \includegraphics[width=\linewidth]{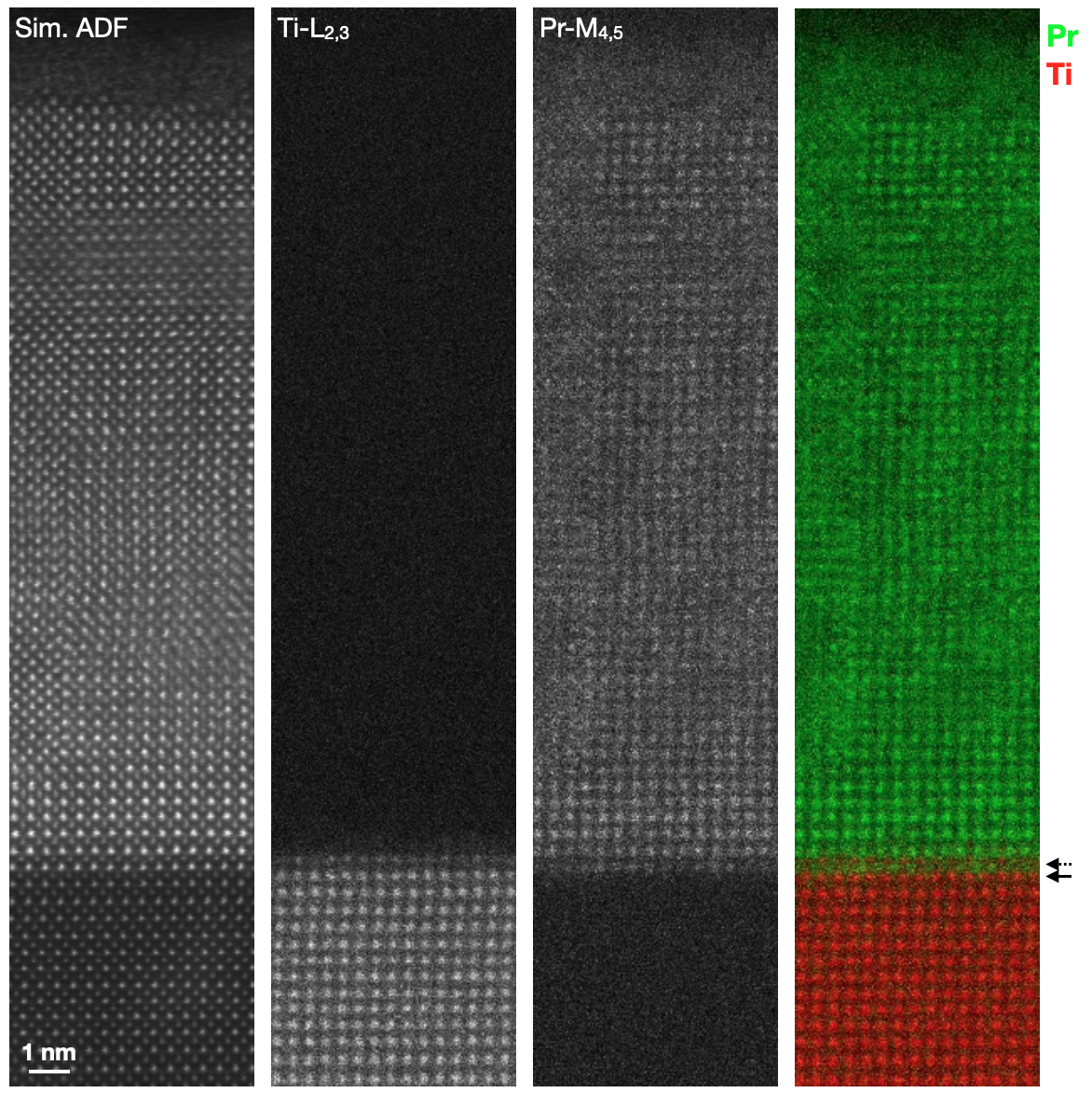}
    \caption{Elemental mapping by STEM-EELS shows the formation of a similar intermediate layer between a Pr$_{0.8}$Sr$_{0.2}$NiO$_2$ film \cite{osada2020superconducting} and SrTiO$_3$ substrate. }
\end{figure}

\begin{figure}
    \centering
        \includegraphics[width=0.75\linewidth]{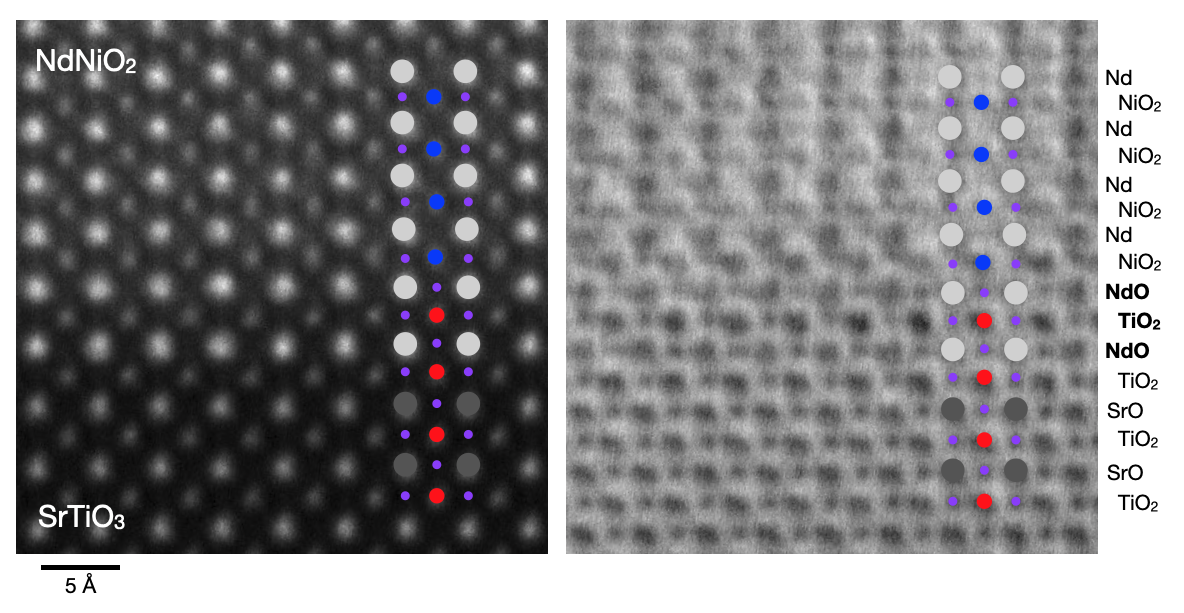}
    \caption{ Annular bright-field (ABF)-STEM imaging reveals oxygen occupancy near the SrTiO$_3$ - NdNiO$_2$ interface which appears consistent with the structure inferred by EELS and DFT.}
\end{figure}

\begin{figure}
    \centering
        \includegraphics[width=0.75\linewidth]{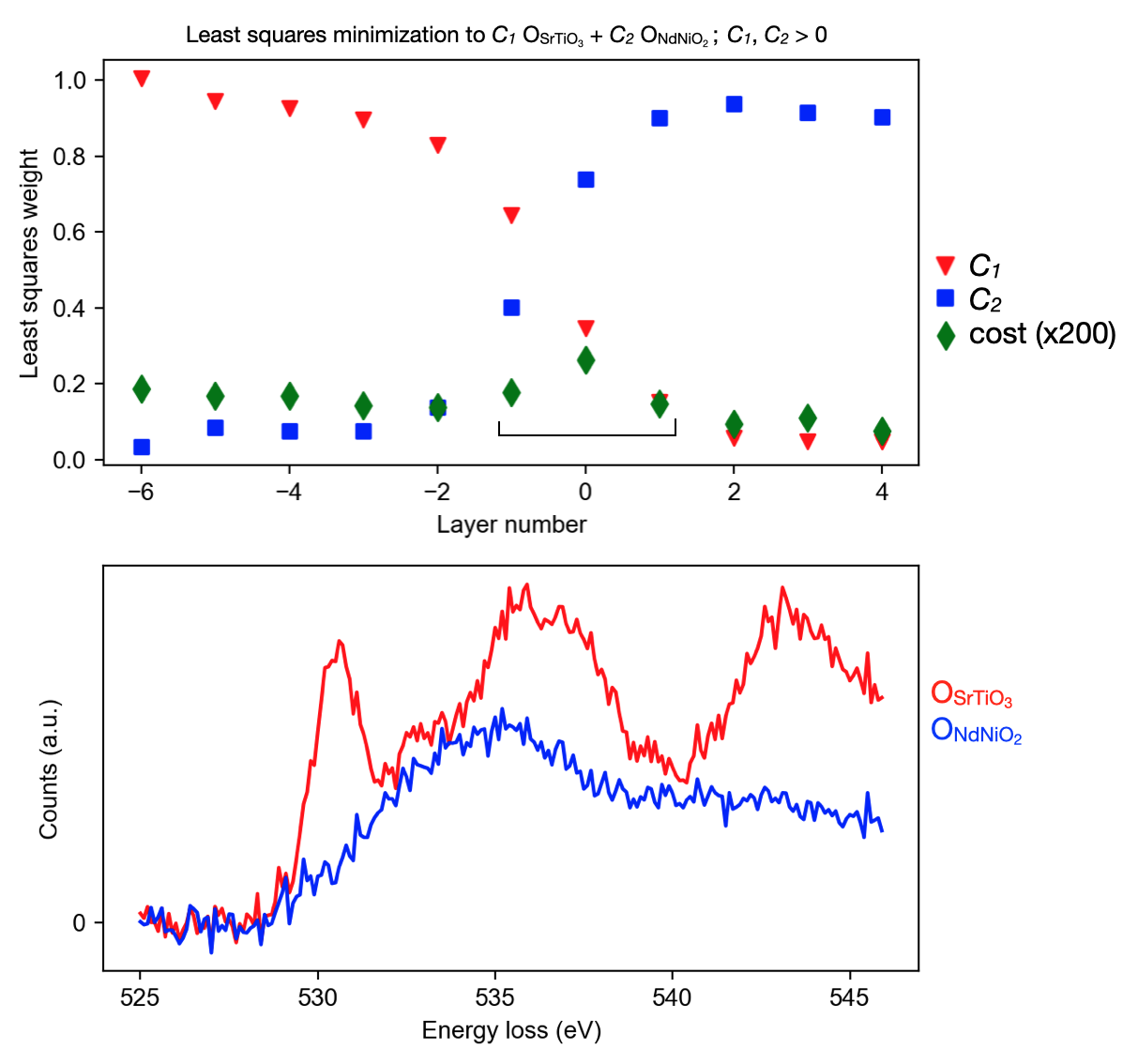}
    \caption{Summary of the least squares fits for each EEL spectrum across the interface for each atomic layer. The calculated weights $A$ and $B$ for the model $A \times \textrm{O}_{SrTiO_3} + B \times \textrm{O}_{NdNiO_2}$, where O$_{SrTiO_3}$ and O$_{NdNiO_2}$ are the reference spectra for SrTiO$_3$ and NdNiO$_2$, respectively, and the cost of each minimization. Cost values are multiplied by 200 for display on the same $y$-axis.}
\end{figure}

\begin{figure}
    \centering
        \includegraphics[width=\linewidth]{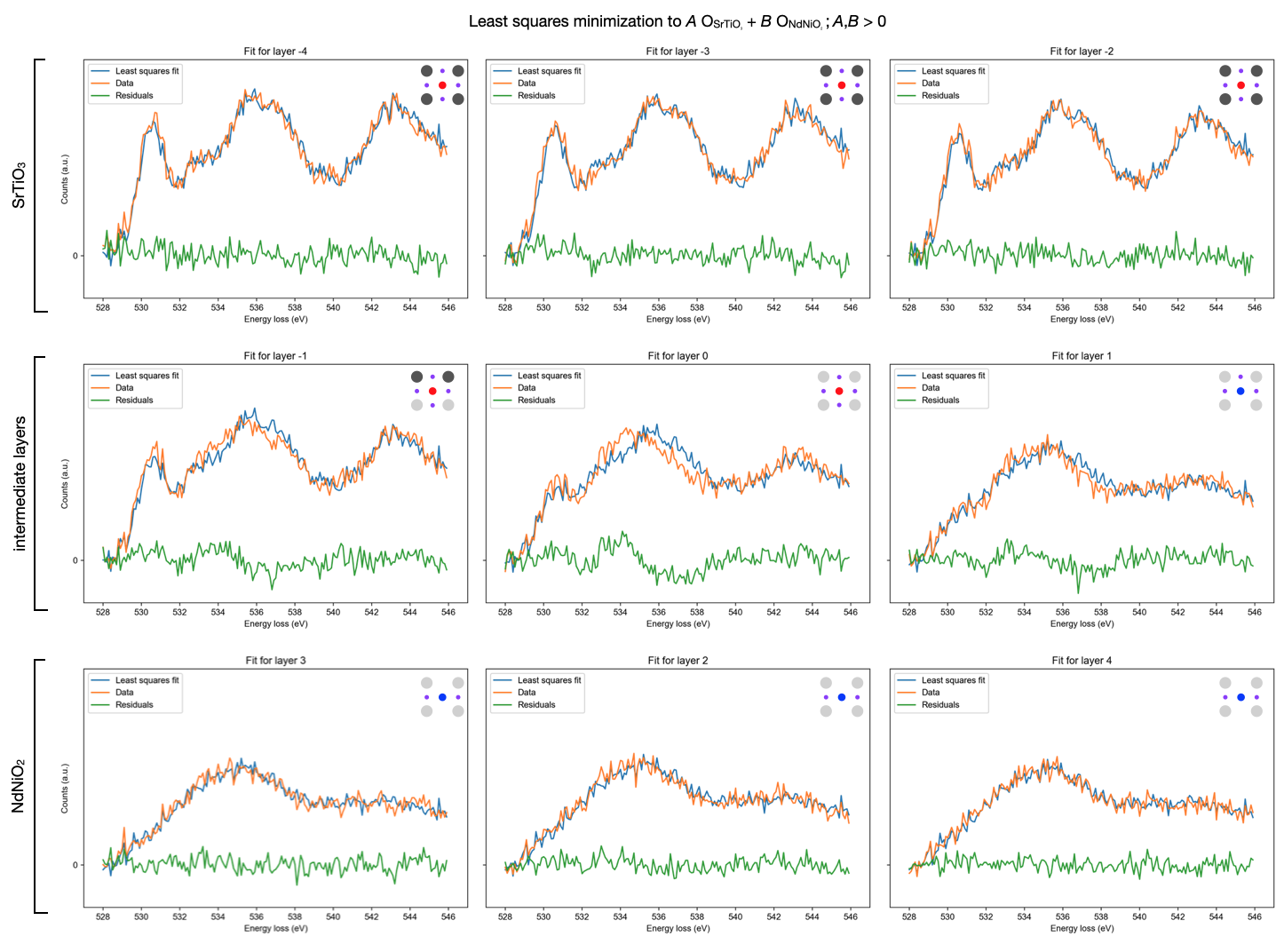}
    \caption{Least squares fits for the EEL spectrum at each atomic layer across the interface as a linear combination following the model $A \times \textrm{O}_{SrTiO_3} + B \times \textrm{O}_{NdNiO_2}$, where O$_{SrTiO_3}$ and O$_{NdNiO_2}$ are the reference spectra for SrTiO$_3$ and NdNiO$_2$, respectively. The weighted contributions $A,B$ of each reference are constrained to be non-negative. The original spectrum (orange) and the result of least squares minimization (blue) are shown, with the residuals (green).  }
\end{figure}

\clearpage

\printbibliography